\documentclass[fleqn,twoside]{article}
\usepackage{espcrc2}

\usepackage{graphicx}
\usepackage[figuresright]{rotating}

\usepackage{amsfonts}
\usepackage{latexsym}

\newcommand{\AmS}{{\protect\the\textfont2
  A\kern-.1667em\lower.5ex\hbox{M}\kern-.125emS}}
\newcommand{\Pp}{P_\pi}
\newcommand{\Fp}{\Phi_\pi}
\newcommand{\Lq}{\Lambda_{\rm QCD}}

\title{The Pion Light-Cone Wave Function $\Fp$ on the lattice: a partonic signal?}

\author{A.~Abada\address{Laboratoire de Physique
		Th\'eorique$^\dagger$ \\
		Universit\'e Paris-Sud, B\^atiment 210, 91405 Orsay Cedex,
		France},  
        Ph.~Boucaud\addressmark,
	G.~Herdoiza\addressmark,
	J.P.~Leroy\addressmark,
	J.~Micheli\addressmark,
	O.~P\`ene\addressmark
	~and~
	J.~Rodr\'{\i}guez--Quintero\address{Dpto. de F\'{\i}sica Aplicada e 
		Ingenier\'{\i}a el\'ectrica \\
		E.P.S. La R\'abida, Universidad de Huelva, 
		21819 Palos de la fra., Spain}
	\thanks{Talk given by Gregorio Herdoiza.
		\hfill \break 
		$^\dagger$Unit\'e Mixte de Recherche du CNRS - UMR 8627.}
        }

\begin{document}

\begin{abstract}

We determine the conditions required to study the pion light-cone wave function
$\Fp$ with a new method: a direct display of the partons constituting the pion.
We present the preliminary results of a lattice computation of $\Fp$ following
this direction. An auxiliary scalar-quark is introduced. The spectroscopy of
its bound states is studied. We observe some indications of a partonic behavior
of the system of this scalar-quark and the anti-quark.

\end{abstract}

\maketitle

\vspace*{-13mm}

\section*{INTRODUCTION}

The light-cone wave function (LCWF), $\Fp(u)$, plays an important role in the
study \cite{Brodsky:1980ny} of pions with large momenta $\Pp$. It is related to the probability of finding in the pion a quark and an
anti-quark moving in the same direction than the pion with momenta $u\Pp$ and
$(1-u)\Pp$ respectively. Within this picture, the two valence quarks are frozen in
a transverse region of size $\Lq$. To extract the partonic information
contained in $\Fp(u)$ a ``kick'' is applied to the pion with a large transfer
momentum $\vec q$. This is for example what has been done in the E791
experiment \cite{E791} for the diffractive dissociation of a pion in two jets.

The LCWF $\Fp$ depends on a momentum scale which is typically $\Pp$. In the
large pion momentum frame, $\Fp$ describes the leading term of the
$\Lq^2/\Pp^2$ expansion of the full pion wave function. It contributes only to
the valence configuration $|q \bar {q} \rangle$ of the Fock space expansion.
$\Fp$ contains all the terms of the $1/\log(\Pp^2/\Lq^2)$ expansion and is
defined by an expression {\it a la} Bethe-Salpeter involving the pion to vacuum
matrix element of a non-local operator

\vspace*{-3mm}

{ \small
\begin{eqnarray} \label{BS}
	&&\hspace*{-7mm}\langle 0| ~\bar{d}(0)
	{\cal{P}}\left[\ \exp(i\int_x^0 \ d\tau_\mu A^\mu)  \right]
	\gamma_{\mu}\gamma_5 u(x)~
	|\pi(\Pp)\rangle_{ \,x^2=0 } \;\cr 
	&&  = -~i~P_{\pi_\mu}~f_\pi \int_0^1 du\, e^{-iuP_\pi\cdot x} 
	\Phi_\pi(u)
\end{eqnarray}
}

\vspace*{-5mm}

The Wilson string in the square bracket ensures the gauge invariance
of the matrix element. $\Fp$ is a non-perturbative function, normalized by the
condition $\int du  \Phi_\pi(u) =1$. 

An important particular case is given by keeping only the dominant term in the
expansion of $\mathcal{O} (1/\log(\Pp^2/\Lq^2))$. It leads to the asymptotic form
of $\Fp$, $\Phi^{\rm as}_\pi(u) = 6 u (1-u)$, which is fully computed by
perturbative QCD\footnotemark[1]\footnotetext[1]{$f_\pi$ being the only non-perturbative
quantity on this limit.}.

On the lattice, the first method to study the LCWF is to examine the moments
${\cal M}_n$ of $\Fp$ \cite{Mom_marti}, ${\cal M}_n = \int_0^1 du \
u^n~\Phi_\pi(u)$. This requires the computation of the matrix elements of
local operators. For high $n$, ${\cal M}_n$ involves operators with high
order derivatives, which are increasingly difficult to discretize and
renormalize. It was therefore proposed in \cite{Dir_marti} to calculate $\Fp$
by studying directly the {\it partonic} constituents of the pion. Following this
idea, we have begun our work by stating the conditions needed to extract
$\Fp$, and then achieved a first and preliminary attempt to isolate a
partonic signal on the lattice \cite{Lcwf}. 

\section{FEASIBILITY CONDITIONS}

\begin{figure}[htb]
\vspace*{-7.5mm}
\includegraphics[width=7.4cm,height=3.5cm]{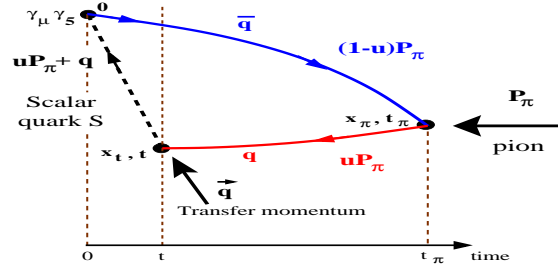}
\vspace*{-9mm}
\caption{\sl \small Three--point Green function computed on the lattice.}
\label{fig_3pt}
\vspace*{-7.5mm}
\end{figure}

The r.h.s of eq.~(\ref{BS}) is computed {\it via} a three-point Green function (see
eq.~(\ref{Fmu}) and fig.~\ref{fig_3pt}). The Wilson string needed to insure
gauge invariance is replaced by a {\it scalar coloured propagator} $S(\vec
{x_t},t;0)$ with the colour quantum numbers of a
quark, but this choice is not unique.

\noindent The active quark is transformed into the scalar-quark at time $t$ and
receives a large momentum transfer $\vec{q}$.

Exhibiting a partonic behavior on the lattice and thus computing directly $\Fp$
requires several conditions which have been derived in \cite{Lcwf} and are 
 listed below without their derivation.

{\it First condition}: large pion momentum, $\Pp \gg \Lq$, to have a
picture of a ``frozen'' pion with a pair of almost colinear valence quarks.
This means that we will have to deal with two scales, $\Lq$ and $\Pp$, on the
lattice. 

{\it Second condition}:  the scalar-quark has to be an energetic parton, $E_s
\gg \Lq$.

{\it Third condition}: the time $t$ must be small to prevent the loss of quantum
coherence between partons and the hadronisation. In particular we need $t \ll (
1 / \Lq  ) \times (  E_s / \Lq  )$: in the pion rest frame, the partonic
behavior lives at a time $\tau \ll 1 / \Lq$; the factor $E_s / \Lq$ acts as a
sort of Lorentz factor. When $t$ becomes large, the partonic system is replaced
by a hadronic state, a bound state made of a quark and a scalar-quark that we
will write $\widetilde \pi$.

{\it Last condition} stemming from finite volume restrictions on the
lattice: in the continuum the momentum fraction $u\in[0,1]$, but due to
momentum discretization on the lattice, $\Pp = 2\pi/L_x \times n_\pi$, we will
only cover a discrete set of values of $u$. Discarding $u = 0~{\rm or}~1$,
because partons at rest is meaningless, we deal with $n_\pi \geq 2$. If $n_\pi
= 2$, we are left with the only value $u = 1 / 2$. 

With all these conditions verified, we compute the three-point Green function
$F^\mu$:

\vspace*{-5mm}

{ \small 
\begin{eqnarray} \label{Fmu}
\hspace*{-1mm}
F^{\mu}(\vec {p_\pi},\vec q;t)
\hspace*{-2.5mm} &\equiv&
\hspace*{-2mm} \int d^3x_\pi d^3x_t\,  e^{-i\vec {p_\pi}\cdot \vec {x_\pi}}\, 
e^{-i\vec q\cdot \vec{x_t}}\, e^{E_\pi (t_\pi - t)} \; \nonumber \\
&&
\hspace*{-14mm} \times \langle  0|   P_5(\vec {x_\pi}, t_\pi)~ u(\vec {x_t},t) 
~  S(\vec {x_t},t;0) ~ \gamma_{\mu} \gamma_5 ~  \bar d(0)   |0 \rangle  \;
\nonumber \\
\hspace*{-3.5mm} &\propto& \hspace*{-2mm} p_{\pi_{\mu}} \, f_\pi
\, \sum_{u_i} \, \frac{ e^{- ( E_s + (1-u_i) E_\pi )\,t}}{2 E_s(u_i)} \,
\Phi_\pi(u_i)
\hspace*{6mm} 
\end{eqnarray}
}

\vspace*{-3.5mm}

The matrix element now contains the interpolating field of the pion $P_5$ at the
fixed time $t_\pi$ and the scalar propagator. We have also included the inverse
pion propagator $e^{E_\pi (t_\pi - t)}$ and required the time interval $t_\pi -
t$ to be large in order to have, at the small times $t$, an on shell pion. In
the r.h.s of eq.~(\ref{Fmu}) we have replaced the integral over $u$ by the sum
because of the discretization constraint. It is important to note that the
fraction which multiplies $\Fp$ can be understood within {\it a partonic
picture} as the product of the inverse scalar propagator $e^{- E_s\,t}/2 E_s$
times the inverse anti-quark propagator $e^{-(1-u_i) E_\pi\,t}$. 

On the lattice we have tried to fulfill as much as possible the
restrictive conditions just presented. This seriously constrain the
simulation and turn the search of a signal into a delicate task. 

\begin{figure*}[t!]
\includegraphics[width=7.5cm,height=3.4cm]{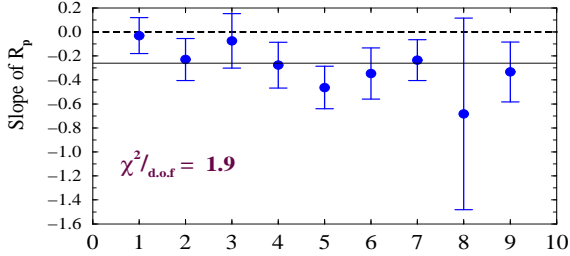}
\hspace*{+6mm} 
\includegraphics[width=7.5cm,height=3.4cm]{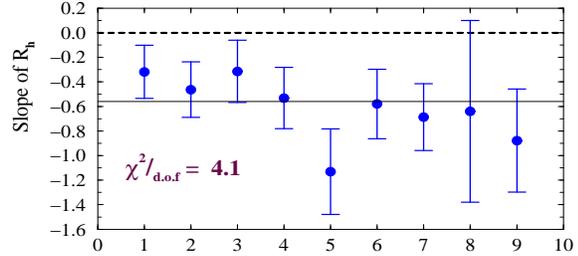} \\
\vspace*{-13mm}
\caption{\sl \small Slopes of the partonic ratio $R_p$ and of the hadronic ratio
$R_h$ in the small time $t$ interval $[0,4]$. The horizontal axis displays the
set of momenta $\vec {q}$ labelled with an arbitrary index.}
\label{Rat}
\vspace*{-4mm}
\end{figure*}  

\section{LATTICE RESULTS}

The simulation was performed on a $16^3 \times 40$ lattice ($\beta=6.0$, $100$
configurations) in the  quenched approximation. We selected light parton masses
($\kappa_q\in\{ 0.1333, 0.1339 \}$ for the quarks and $\kappa_s\in\{ 0.1428,
0.1430, 0.1431 \}$ for the scalar) and the following values for $\vec {\Pp}$,
$\vec {\Pp}\in\{ (0,0,0); (1,0,0); (1,1,0); (2,0,0) \}$ in units of $2\pi /
L_x$. A large set of transfer momenta $\vec {q}$ was chosen.

As we have introduced in eq.~(\ref{Fmu}) a non physical object, the
scalar-quark, the first step of the analysis was to test its hadronic behavior.
We measured therefore the two-point Green functions for the $\widetilde \pi$
and the scalar-scalar states. These two-point functions show the characteristic
exponential signal of hadrons and verify the Einstein spectral law. Interestingly, the spectroscopy of the scalar reveals
that these non-conventional bound states behave as real hadrons.

We will now consider the three-point Green function $F^{0}(\vec {\Pp},\vec
q;t)$, eq.~(\ref{Fmu}). As already justified, we use the momentum
$\vec{\Pp}\,=\,(2,0,0)\times2\pi / L_x$ in order to have $u = 1 / 2$ although
large momenta bring up some noise in a lattice signal. 

We will try to give a partonic interpretation of the data in the
short time $t$ region by building the following ratios

\vspace*{-5mm}

{ \small 
\begin{eqnarray*}
R_p &=& \hspace*{+5mm} \frac{ F^0(\vec p_{\pi},\vec q ;t)  }{ \left[
 \, p_{\pi}^0 f_\pi \,\frac{ e^{- ( E_\pi/2 + E_s)\,t} }{ 2 E_s } \,
\Phi_\pi(1/2) \right] } \\
R_h &=& \hspace*{+18mm} \frac{ F^0(\vec p_{\pi},\vec q ;t) }{\left[ e^{- E_{\widetilde\pi}
\,t}\right]} 
\end{eqnarray*}
}

\vspace*{-4mm}

We will look for plateaus in time of these quantities for the whole set of
momenta $\vec q$. A plateau for $R_p$ ($R_h$) would be a sign of a partonic
(hadronic) behavior of the antiquark--scalar system. In fig.~\ref{Rat}, we
plot the {\it slopes} of the ratios $R_p$ and $R_h$ in the short time region $t
\in [0,4]$, for our set of momenta $\vec q$.  A value
of the slope close to zero suggests a plateau in time. By comparing both
plots, we observe that the partonic slopes are closer to zero than the
hadronic ones. According to our data, the partonic interpretation is somehow
favoured: $\chi^2 / {\rm d.o.f}=1.9$ for a vanishing slope compared to $4.1$
in the hadronic case. Nevertheless, we are clearly not yet in a position to
give a fully convincing claim. 

\begin{figure}[htb]
\includegraphics[width=7.4cm,height=3.4cm]{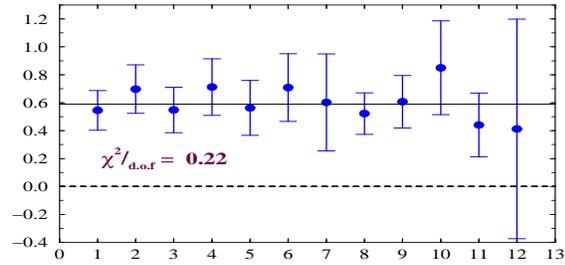} 
\vspace*{-8.5mm}
\caption{\sl\small Values of $F^{0}(\vec {p_\pi}\,=\,(2,0,0),\vec q ; t\,=\,0
)\times E_s(1 / 2)$ for our full set of momenta $\vec q$.}
\label{F0}
\vspace*{-7mm}
\end{figure}

As a cross-check of the previous result, we have examined the product
$F^{0}(\vec {p_\pi}\,=\,(2,0,0),\vec q ; t\,=\,0 )\times E_s(1 / 2)$. Within the
partonic picture, we expect this product to be independent of the transfer
momentum $\vec q$ (see eq.~(\ref{Fmu})). This is indeed what we find when we
plot in fig.~\ref{F0} this product for our full set of momenta $\vec
q$. The data show the expected constancy around the average represented by the
solid horizontal line. 


\section*{CONCLUSION}

\vspace*{-0.5mm}

We wanted to check the feasibility on the lattice of a direct method to determine
non-perturbative objects like the pion light-cone wave function $\Fp$. We have
stated the conditions for this study and we have seen that they induce
restrictive constraints and hierarchies which turn the search of a clear lattice
signal into a difficult task. We have found an interesting spectroscopy of bound
states involving a scalar-quark, and we have a hint of a partonic behavior of the
system of the scalar and the anti-quark by considering the three-point Green
function related to $\Fp$. This signal will continue to be tested in a further
analysis. Work supported by the European Community's Human potential programme under
HPRN-CT-2000-00145 Hadrons/LatticeQCD.

\end{document}